\documentclass[runningheads]{llncs}
\usepackage[T1]{fontenc}
\usepackage{graphicx}
\usepackage{subcaption}
\usepackage{hyperref}
\usepackage{amssymb}
\usepackage{booktabs}
\usepackage{amsmath}
\usepackage{flafter}

\begin{document}

\title{
Validating the Clinical Utility of CineECG 3D Reconstructions through Cross-Modal Feature Attribution
}
\titlerunning{
Validating CineECG via Cross-Modal XAI
}
\author{Karol Dobiczek\inst{1}\thanks{These authors contributed equally to this work.}\orcidID{0009-0001-0906-2393} \and
Maciej Mozolewski\inst{1}$^\star$\orcidID{0000-0003-4227-3894} \and
Szymon Bobek\inst{1}\orcidID{0000-0002-6350-8405} \and
Michał Szafarczyk\inst{2}\orcidID{0009-0000-2270-5106} \and
Peter van Dam\inst{2}\orcidID{0000-0002-7962-1760} \and
Grzegorz J. Nalepa\inst{1}\orcidID{0000-0002-8182-4225}}
\authorrunning{K. Dobiczek et al.}
\institute{Department of Human-Centered Artificial Intelligence, Institute of Applied Computer Science, Jagiellonian University, Krakow, Poland\\
\email{karol.dobiczek@doctoral.uj.edu.pl},
\email{\{m.mozolewski,szymon.bobek,grzegorz.j.nalepa\}@uj.edu.pl} \and
Faculty of Medicine, Jagiellonian University Medical College, Krakow, Poland\\
\email{michal.szafarczyk@student.uj.edu.pl}, \email{peter.dam@uj.edu.pl}}
\maketitle

\begin{abstract}
    Deep learning models for 12-lead electrocardiogram (ECG) analysis achieve high diagnostic performance but lack the intuitive interpretability required for clinical integration.
    Standard feature attribution methods are limited by the inherent difficulty in mapping abstract waveform fluctuations to physical anatomical pathologies.
    To resolve this, we propose a cross-modal method that projects feature attributions from high-performance 12-lead ECG models onto the CineECG 3D anatomical space.
    Our study reveals that while models trained directly on CineECG signals suffer from reduced accuracy and incoherent attributions, the proposed mapping mechanism effectively recovers clinically relevant feature rankings.
    Validated against a ground-truth dataset of 20 cases annotated by domain experts, the mapped explanations yield a Dice score of 0.56, significantly outperforming the 0.47 baseline of standard 12-lead attributions.
    These findings indicate that cross-modal averaging mapping effectively filters attribution instability and improves the localization of pathological features, combining the diagnostic expressiveness of standard ECG with the intuitive clarity of anatomical visualization.
\end{abstract}

\section{Introduction} \label{sec:introduction}

\begin{figure}[t]
    \centering
    \includegraphics[width=0.8\linewidth]{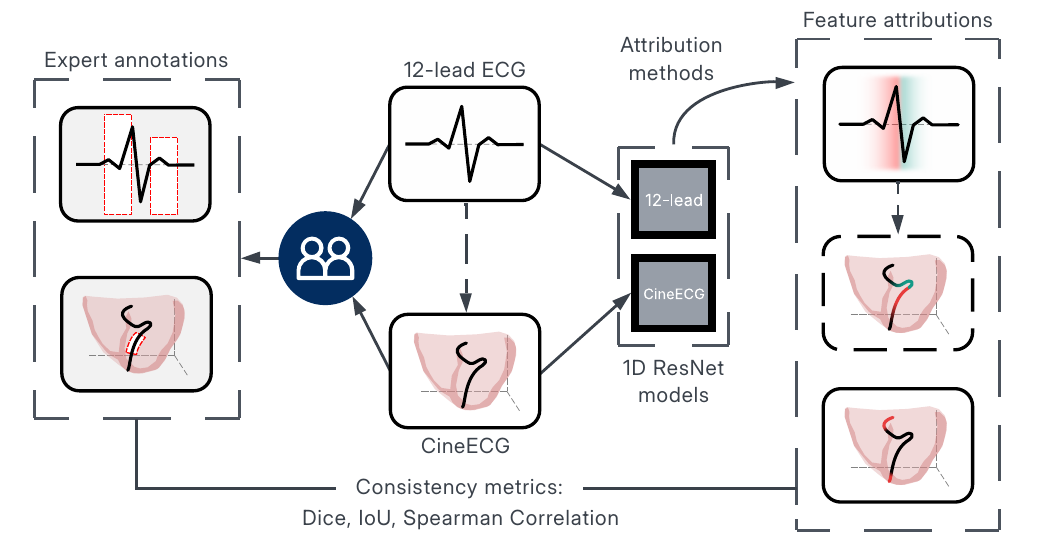}
    \vspace{-10pt}
    \caption{Separate models are trained on ECG and CineECG data. Model feature attributions are compared to ground-truth annotations produced by experts.}
    \label{fig:general_diagram}
    \vspace{-10pt}
\end{figure}

The 12-lead electrocardiogram (ECG) remains the primary diagnostic tool for assessing cardiovascular electrical activity.
However, the clinical interpretation of these temporal signals does not directly associate waveforms with specific cardiac structures, and requires substantial expertise and pattern recognition skills.
This cognitive burden can lead to inter-clinician variability and diagnostic uncertainty~\cite{wood_exploring_2014}.
To address this limitation in data representation, the CineECG method reconstructs the propagation of the cardiac impulse directly in a 3D space~\cite{van_dam_novel_2020}.
By providing a direct anatomical frame of reference, CineECG fundamentally enhances the clinical interpretability of the signal.

In parallel to advancements in clinical visualization, deep learning models trained on standard 12-lead ECGs have achieved exceptional diagnostic accuracy~\cite{aziz_ecg-based_2021}.
The clinical deployment of these opaque systems critically depends on the explainability of their predictions to ensure safety and foster physician trust \cite{taleban_explainable_2026}.
Explainable AI (XAI) methods such as feature attributions aim to provide this transparency. By calculating and highlighting the salient segments of the input such as the raw 12-lead ECG signal they produce explanations to the models' decisions.
However, mapping these raw signals to physical anatomical pathologies remains clinically challenging.
As demonstrated by Patlatzoglou et al.~\cite{patlatzoglou_cost_2025}, standard post-hoc techniques often produce approximations of the model's true behavior that are ambiguous, unreliable, and inconsistent.
Furthermore, van de Leur et al.~\cite{van_de_leur_improving_2022} noted that such heatmap-based methods hold limited explainable value to identify the actual morphological changes in ECG driving a prediction.
Consequently, it is imperative to develop explainability methods capable of translating the abstract temporal focus of the model into the spatial reasoning required by the cardiologist.

In this work, we propose a cross-modal explainability method that combines the predictive power of 12-lead models with the anatomical clarity of CineECG.
We demonstrate that training neural networks exclusively on the CineECG 3D spatial trajectories degrades both predictive performance of the models and the coherence of generated explanations.
Instead, our approach visualized in Figure~\ref{fig:general_diagram} utilizes the standard 12-lead architecture as a primary feature extractor and through an XAI algorithm, derives temporal attributions that are subsequently projected onto the CineECG ventricular model.
Evaluated against a ground-truth dataset of expert annotations, this cross-modal mapping yields a significantly higher alignment with clinical assessment compared to standard ECG baselines.
This approach effectively bridges the gap between the algorithmic complexity of deep learning and the clinical reasoning of medical professionals.

The remainder of this paper is organized as follows.
Section~\ref{sec:background} outlines the research background, followed by the proposed cross-modal mapping methodology in Section~\ref{sec:method}.
Experimental results are presented in Section~\ref{sec:results}, with clinical implications and limitations discussed in Section~\ref{sec:discussion}.
Section~\ref{sec:conclusions} concludes the study.

\section{Background} \label{sec:background}

Feature attribution methodologies are XAI methods that quantify the relative contribution of input features to a model's specific prediction, facilitating a granular understanding of the underlying decision-making process.
Due to the ever-increasing use of ML models for clinical ECG tasks such as diagnoses or classifications, previous works use feature attributions as means of increasing trust in the model predictions or as an aid in diagnosis interpretation~\cite{taleban_explainable_2026}. The model agnostic SHAP and LIME methods are commonly used \cite{majhi_explainable_2024} along with gradient-based approaches such as GradCAM and IG~\cite{ojha_exploring_2024,m_explainable_2023}. 
As a relatively novel method, CineECG has mainly been used as data source for statistical~\cite{boonstra_novel_2021} and classical ML classification models~\cite{mortada_quantifying_2025}. To our best knowledge this work is the first to explore AI explainability in conjunction with CineECG.

In this study, we evaluate and compare two distinct families of attribution techniques: those based on local surrogate methods and those utilizing the model's gradient.
The first category includes \emph{Local Interpretable Model-agnostic Explanations} (LIME)~\cite{ribeiro_why_2016}, which approximates the decision boundary by fitting a linear surrogate model to localized perturbations of the original input.
Within the same family, \emph{KernelSHAP}~\cite{lundberg_unified_2017} leverages cooperative game theory to derive \emph{Shapley values} through a weighted linear regression, estimating the expected change in prediction associated with each feature.

The second category comprises gradient-based approaches that leverage the differentiability of deep neural networks to assign importance.
\emph{GradientSHAP} approximates Shapley values by computing the expectation of gradients across a distribution of randomly sampled baselines, scaling the result by the difference between the original input and the reference point.
Similarly, \emph{Integrated Gradients} (IG)~\cite{sundararajan_axiomatic_2017} addresses the saturation limitations of standard gradients by computing a path integral from a baseline input to the target sample.
This approach is mathematically grounded in the axioms of sensitivity and implementation invariance, ensuring that the total attribution sum accounts for the entire change in the model's output relative to the neutral baseline.

\section{Method} \label{sec:method}

\subsection{Data Selection and Preparation} \label{subsec:dataset}
We utilized a synchronized subset of the PTB-XL dataset~\cite{wagner_ptb-xl_2022} comprising 8,000 records equally balanced between \emph{Normal} and \emph{Abnormal} classes.
To ensure ground-truth reliability, the selected records underwent clinical validation by cardiologists, with any identified diagnostic misclassifications manually corrected.
The integration of standard 12-lead ECG signals with their corresponding CineECG counterparts ensured precise temporal alignment across both modalities.
Each recording was truncated to a standardized 400\,ms window to focus on the QRS-T complex, preserving the essential morphological information for diagnostic analysis.
The CineECG signals were derived directly from these pre-processed signal fragments to maintain strict consistency between data representations.
The main corpus was partitioned into training, validation, and test sets following a 60/20/20\% ratio using a stratified shuffle split at the patient level.

To ensure the integrity of the explainability analysis, we curated an additional evaluation cohort of 20 independent cases from a separate patient pool.
This subset remained entirely disjoint from the model development phase, providing a baseline for the expert-grounded alignment analysis without the risk of data leakage.
Furthermore, we established a physiological reference group of 150 normal samples to serve as the baseline distribution for the \emph{SHAP} attribution estimates.
This isolation ensures that the resulting feature attributions reflect the model's underlying generalization capability rather than the memorization of training instances.

\subsection{Classification Architecture and Training Protocol} \label{subsec:model_architecture}
\begin{figure}[htbp]
    \centering
    \includegraphics[width=\linewidth]{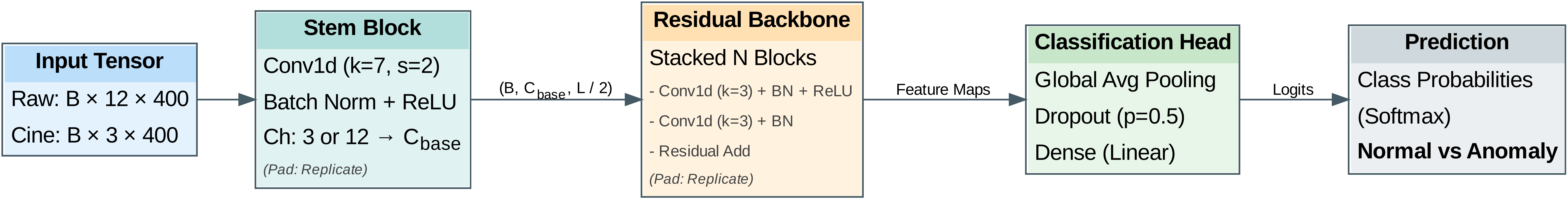}
    \caption{XAI-optimized 1D-ResNet architecture designed for variable-length ECG classification.}
    \label{fig:resnet_arch}
\end{figure}

The diagnostic classification relies on a customized 1D-ResNet network implemented in \texttt{PyTorch} (v2.6.0), performing a two class (normal, abnormal) classification, and specifically engineered to maintain strict differentiability for gradient-based explanations.
As illustrated in Figure~\ref{fig:resnet_arch}, the network initializes with a convolutional stem utilizing strided convolution for downsampling rather than standard pooling operations.
This architectural decision was made to prevent shattered gradients within the subsequent attribution maps.
The core feature extractor comprises stacked residual blocks incorporating replication padding to mitigate boundary discontinuities and prevent artifactual importance scores at the signal edges.
A Global Average Pooling layer aggregates the temporal features, rendering the classification head agnostic to the input sequence length and establishing a unified architecture for both the multi-lead raw ECG and the 3-channel CineECG signals.
The resulting latent vector is regularized via dropout operations before the final dense projection.

To maximize the generalization capability across both data modalities, we conducted a systematic hyperparameter search utilizing the \texttt{Optuna} framework (v4.6.0).
The optimization process employed the Tree-structured Parzen Estimator to traverse the search space over 130 trials, identifying optimal structural and learning parameters.
Final model training was executed utilizing the Adam optimizer to minimize the Cross-Entropy loss function.
We incorporated an early stopping mechanism monitored on an additional 15\% validation split held-out from the training set to mitigate overfitting.
Evaluation on the independent test set established the performance metrics of both selected architectures.
The 12-lead ECG model achieved an accuracy of 83.06\% with a macro F1-score of 0.8301.
Operating on the reduced-dimensionality spatial signal, the CineECG model attained an accuracy of 79.06\% and a macro F1-score of 0.7902.

\subsection{Computational Implementation and Parameters} \label{subsec:attribution_methods}
The computational pipeline for generating feature attributions was implemented using \texttt{shap} (v0.50.0), \texttt{lime} (v0.2.0.1), and \texttt{captum} (v0.8.0).
Gradient SHAP utilized a reference distribution of 150 normal subjects.
Kernel SHAP processed the multidimensional time-series as flattened tabular vectors with a sampling budget of 100 evaluations per instance.
Similarly, the approximated local decision boundaries by training a linear surrogate on 500 perturbed samples, with individual time-steps treated as discretized features.
Finally, Integrated Gradients used a 50-step path integral and averaged attributions across 40 Gaussian noise-injected samples with a standard deviation of $\sigma = 0.15$.

\subsection{Cross-Modal Attribution Mapping} \label{subsec:mapping_method}
To bridge the gap between model performance and interpretability, we propose a cross-modal mapping method.
This method enables the transfer of feature importance scores derived from the high-performance 12-lead ECG model to the anatomical CineECG modality.
We first convert the raw attribution maps into a unified bipolar importance profile $\Phi$ according to Equation~\ref{eq:bipolar}, where positive magnitudes indicate evidence supporting abnormal classification.
\begin{equation}
    \Phi_{c,t} = \frac{A_{c,t}^{(1)} - A_{c,t}^{(0)}}{2}
    \label{eq:bipolar}
\end{equation}
Here, $A_{c,t}^{(k)}$ denotes the raw attribution for channel $c$ at the time-step $t$ with respect to class $k$ (where $k=1$ represents the abnormal class and $k=0$ the normal class), and $\Phi_{c,t}$ constitutes the resulting standardized score preserving the dimensionality of the channel.

Subsequently, we aggregate the standardized 12-lead attributions into a global temporal vector and normalize the result to the unit interval $[-1, 1]$ as defined in Equation~\ref{eq:projection}.
\begin{equation}
    \Phi_{d,t} = \frac{\sum_{l=1}^{L} \Phi_{l,t}^{12-lead}}{\max_{t} \left| \sum_{l=1}^{L} \Phi_{l,t}^{12-lead} \right|}
    \label{eq:projection}
\end{equation}
In this formulation, the same aggregated scalar value is {uniformly applied to all spatial dimensions} $d \in \{x,y,z\}$ to ensure consistent trajectory coloring at each time-step $t$. Here, $L=12$ refers to the total number of leads in the 12-lead ECG signal, and $\Phi_{l,t}^{12-lead}$ denotes the standardized attribution of the $l$-th lead.

\subsection{Expert Annotation Platform and Protocol} \label{subsec:expert_annotations}
To establish a reliable ground truth for the alignment metrics, we curated a stratified validation subset of 20 PTB-XL records, equally balanced between normal and pathological cases, distinct from the model development subsets.
Independent assessments were conducted using a custom web-based evaluation platform~\footnote{Expert feedback collection platform: \url{https://wiki.iis.uj.edu.pl/cmuj/}} engineered to enforce a strictly blind review process.
This dedicated interface presented the standard 12-lead ECG on a calibrated clinical grid alongside an interactive 3D CineECG trajectory, allowing cardiologists to freely rotate the anatomical model and inspect the electrical activation loop.
By explicitly withholding both the original diagnostic labels and the model's predictions, the environment ensured that the resulting annotations reflected unbiased clinical judgment. 
The experts annotated the signals by leaving their diagnosis and specifying the salient segments in a textual form.
The granular spatiotemporal boundaries recorded through this system were subsequently converted into the binary ground-truth masks utilized for the quantitative XAI alignment analysis.

% \subsection{Expert Annotations and Ground-Truth Masks} \label{subsec:expert_annotations}
To establish a reliable baseline for evaluating attribution validity, we conducted a blind study involving two domain experts.
The experts were provided with a randomly selected validation subset containing synchronized 12-lead ECG and CineECG recordings, alongside the patient's heart rate, while the original diagnostic labels were intentionally redacted.
Although the standard 12-lead ECG and the interactive 3D CineECG were presented concurrently in a side-by-side interface, experts annotated each modality independently.
The annotation protocol required experts to mark these segments using predefined minimal windows of $25$ ms within the QRS complex region ($0-150$ ms) and $50$ ms for the remainder of the cardiac cycle ($150-400$ ms).
In cases where physicians identified instantaneous points of interest, we expanded these specific timestamps by a tolerance window of $\pm 10$ ms to account for minor physiological variations.
These structured temporal annotations were subsequently converted into binary segmentation masks to serve as the computational ground truth.

\subsection{Quantifying Expert-Model Agreement} \label{subsec:quantifying_expert_model_agreement}
Following established methodologies for evaluating attribution alignment~\cite{singh_agreement_2020}, we assessed the agreement between model explanations and expert annotations using a combination of spatial overlap and rank correlation metrics.
Specifically, we used the Intersection over Union (IoU) and the Dice coefficient~\cite{Srensen1948AMO} to measure the spatial similarity of activated regions, alongside Spearman's rank correlation~\cite{spearman_proof_1904} to evaluate the monotonic relationship between raw attribution values and expert derived ground truth.
To prevent unannotated background channels from skewing the evaluation, all metrics were computed on a per-sample basis exclusively within the specific leads identified by the expert.

Transforming the continuous attribution arrays into binary masks for the IoU and Dice calculations required establishing an activation boundary.
To ensure a fair comparison across the two modalities, we optimized the thresholds in the range $[0, 1]$ with a step size of $0.01$ and ultimately select the value that maximized the respective Dice coefficient.
This optimal threshold concurrently serves as an intrinsic indicator of attribution quality, as values approaching zero suggest a failure to isolate focal salient features, effectively treating the entire signal sequence as relevant.

The raw attribution $\Phi$ inherently constitutes a bipolar signal, where positive magnitudes provide evidence for the abnormal class and negative magnitudes support the normal class.
To properly align this output with the binary expert masks for healthy patients, we inverted the attribution sign ($\Phi' = -\Phi$) for cases explicitly diagnosed as normal, ensuring that features supporting the ground-truth class mapped to positive values.
Subsequently, we processed these aligned signals through three distinct transformation strategies to define the final importance map $M_{attr}$.
\emph{Post-processing} strategies included a \texttt{Positive} variant that considered only evidence explicitly supporting the target class ($M_{attr} = \max(0, \Phi')$), an \texttt{Absolute} variant that treated both positive and negative contributions as salient features ($M_{attr} = |\Phi'|$), and a \texttt{Scaled} variant utilizing min-max normalization to preserve the relative continuous feature ranking.

\section{Results} \label{sec:results}
% \subsection{Expert Evaluation Platform} \label{subsec:expert_evaluation}
% To establish a reliable ground truth for the alignment metrics, we curated a stratified validation subset of 20 PTB-XL records, equally balanced between normal and pathological cases and entirely distinct from the model development subsets.
% The independent assessments were conducted utilizing a custom web-based evaluation platform~\footnote{Expert feedback collection platform: \url{https://wiki.iis.uj.edu.pl/cmuj/}} engineered to enforce a strictly blinded review process.
% This dedicated interface presented the standard 12-lead ECG on a calibrated clinical grid alongside an interactive 3D CineECG trajectory, allowing cardiologists to freely rotate the anatomical model and inspect the electrical activation loop.
% By explicitly withholding both the original diagnostic labels and the model's predictions, the environment ensured that the resulting annotations reflected unbiased clinical judgment. 
% The experts annotated the signals by leaving their diagnosis along with specifying the salient segments in a textual form.
% The granular spatiotemporal boundaries recorded through this system were subsequently converted into the binary ground-truth masks utilized for the quantitative XAI alignment analysis.

\subsection{Analysis of Experts vs XAI Agreement} \label{subsec:stat_analysis}
To identify the most effective interpretability configuration, we established a comprehensive \emph{optimization pool} encompassing all experimental variants.
Each annotated clinical case was evaluated across multiple combinatorial dimensions, including the chosen attribution method (Gradient/Kernel SHAP, LIME, and IG), the post-processing strategy applied to the importance signals (positive, absolute, and scaled), and the primary data representation (raw 12-lead ECG, direct CineECG, and mapped 12-lead to CineECG).
For the CineECG analysis, this expansion additionally incorporated both natively computed attributions for CineECG model and those projected from the 12-lead domain.

Within each evaluated context, the method identified the specific configuration of parameters that maximized the mean Dice coefficient across the patient cohort.
Subsequent comparative analyses were restricted to these optimized subsets, ensuring that each interpretability tool was evaluated at its peak performance rather than against a suboptimal baseline.

The explanation alignment metrics were subsequently estimated on these optimized subsets using a non-parametric bootstrapping technique implemented with the \texttt{scipy} (v1.17.0) and \texttt{scikit-learn} (v1.7.2) libraries.
We executed 2000 resampling iterations for each cohort to compute 95\% confidence intervals via the Bias-Corrected and Accelerated (BCa) method.

\subsection{CineECG vs. 12-Lead ECG Explanation Alignment} \label{subsec:bootstrap_best}
The Integrated Gradients method emerged as the superior attribution technique for both signal representations, though optimal post-processing strategies diverged.
Specifically, the CineECG signal achieved maximum alignment utilizing mapped 12-lead attributions with \emph{Scaled} normalization, whereas the 12-lead baseline peaked using the \emph{Absolute} magnitude.
Table~\ref{tab:sota_performance} presents the bootstrap-estimated alignment metrics for these optimized configurations.
The spatial mapping onto the CineECG trajectory yielded a higher mean Dice coefficient of 0.56 compared to the 0.47 baseline established by the raw 12-lead signal.
A similar differential was observed in the Intersection over Union and Spearman correlation scores, consistently favoring the CineECG format.

\begin{table}[htb!]
    \vspace{-15pt}
    \centering
    \caption{Explanation alignment metrics comparing the optimised performance of CineECG and 12-Lead ECG.}
    \label{tab:sota_performance}
    \resizebox{0.9\linewidth}{!}{
    \begin{tabular}{lccc}
        \toprule
        \textbf{Modality} & \textbf{Dice Score} & \textbf{IoU Score} & \textbf{Spearman Cor.} \\
        \midrule
        CineECG Signal & 0.56 (0.51–0.62) & 0.40 (0.35–0.47) & 0.22 (0.11–0.29) \\
        12-Lead Signal & 0.47 (0.44–0.52) & 0.32 (0.29–0.37) & 0.18 (0.12–0.23) \\
        \bottomrule
        \multicolumn{4}{l}{\footnotesize \textit{Note:} Values represent bootstrap-estimated Means (95\% BCa Confidence Interval).}
    \end{tabular}
    }
    \vspace{-10pt}
\end{table}

Further stratification reveals that explanation alignment with expert annotations was notably higher in pathological cases with a mean Dice score of 0.59, compared to 0.53 for the normal subgroup for the CineECG modality.
The 12-lead signal exhibited a comparable but less pronounced trend.
Attribution quality displayed an inverse relationship for diagnostic agreement with model predictions. Paradoxically, the CineECG modality achieved a higher Dice score (0.57) for diverging diagnoses, outperforming agreeing cases (0.50).
This suggests the feature attribution mechanism successfully localizes pathological patterns even when the final classification head fails.
The raw 12-lead signal showed minimal sensitivity to this factor, with statistically indistinguishable performance between agreement states.
Table~\ref{tab:diagnosis_agreement_stability} aggregates these metrics for detailed reference.

\begin{table}[ht!]
    \vspace{-15pt}
    \centering
    \caption{Impact of Diagnosis and Agreement status on explanation alignment metrics.}
    \label{tab:diagnosis_agreement_stability}
    \resizebox{0.9\linewidth}{!}{
    \begin{tabular}{llccc}
        \toprule
        \textbf{Modality} & \textbf{Condition} & \textbf{Dice Score} & \textbf{IoU Score} & \textbf{Spearman Cor.} \\
        \midrule
        \multicolumn{5}{l}{\textit{Stratification by Diagnosis}} \\
        CineECG Signal & Abnormal & 0.59 (0.48–0.71) & 0.44 (0.33–0.57) & 0.19 (0.02–0.33) \\
                       & Normal   & 0.53 (0.49–0.55) & 0.36 (0.33–0.38) & 0.24 (0.09–0.31) \\
        12-Lead Signal & Abnormal & 0.49 (0.39–0.60) & 0.35 (0.26–0.46) & 0.17 (0.01–0.26) \\
                       & Normal   & 0.46 (0.44–0.48) & 0.30 (0.28–0.32) & 0.19 (0.12–0.23) \\
        \midrule
        \multicolumn{5}{l}{\textit{Stratification by Agreement}} \\
        CineECG Signal & Agreement  & 0.50 (0.47–0.52) & 0.34 (0.31–0.35) & 0.07 (-0.31–0.27) \\
                       & Disagreement & 0.57 (0.50–0.63) & 0.41 (0.35–0.49) & 0.24 (0.14–0.31) \\
        12-Lead Signal & Agreement  & 0.47 (0.42–0.50) & 0.31 (0.27–0.34) & 0.19 (0.03–0.30) \\
                       & Disagreement & 0.48 (0.43–0.52) & 0.32 (0.29–0.38) & 0.18 (0.10–0.23) \\
        \bottomrule
        \multicolumn{5}{l}{\footnotesize \textit{Note:} Values represent bootstrap-estimated Means (95\% BCa Confidence Interval).}
    \end{tabular}
    }
    \vspace{-10pt}
\end{table}

\subsection{Comparative Analysis of Attribution Methods} \label{subsec:method_performance}
Optimal post-processing strategies depended on both the signal modality and the attribution method.
All techniques evaluated achieved peak performance on the CineECG signal using \emph{Scaled} normalization.
Conversely, the 12-lead baseline required \emph{Absolute} magnitudes for gradient-based methods (GradSHAP, IG) and \emph{Scaled} normalization for perturbation-based methods (Kernel SHAP, LIME).

Table~\ref{tab:benchmark_1_methods} details the explanation alignment metrics for each attribution method under its optimal post-processing configuration.
IG emerged as the most effective technique across both domains, achieving a mean Dice score of 0.56 for the mapped CineECG and 0.47 for the standard 12-lead signal.
Gradient SHAP followed closely as the second most accurate method, particularly in the CineECG modality where it significantly outperformed its 12-lead counterpart.
Conversely, the perturbation-based methods, encompassing Kernel SHAP and LIME, yielded consistently lower spatial overlap scores across all test conditions with the Spearman correlation analysis highlighting a sharp distinction between the method families.

\begin{table}[bt!]

    \vspace{-15pt}
    \centering
    \caption{Attribution Methods Performance under optimized post-processing.}
    \label{tab:benchmark_1_methods}
    \resizebox{0.9\linewidth}{!}{
    \begin{tabular}{llccc}
        \toprule
        \textbf{Modality} & \textbf{Attribution} & \textbf{Dice Score} & \textbf{IoU Score} & \textbf{Spearman Cor.} \\
        \midrule
        CineECG  & Gradient SHAP & 0.52 (0.47–0.58) & 0.36 (0.32–0.43) & 0.09 (0.03–0.14) \\
        Signal   & Integrated Grad. & 0.56 (0.51–0.62) & 0.40 (0.35–0.47) & 0.22 (0.11–0.29) \\
                 & Kernel SHAP & 0.46 (0.41–0.53) & 0.31 (0.27–0.39) & $<0.01$ (-0.01–$<0.01$) \\
                 & LIME & 0.46 (0.42–0.54) & 0.31 (0.27–0.40) & 0.01 (-0.01–0.04) \\
        \midrule
        12-Lead  & Gradient SHAP & 0.47 (0.44–0.52) & 0.32 (0.29–0.37) & 0.17 (0.10–0.23) \\
        Signal   & Integrated Grad. & 0.47 (0.43–0.51) & 0.32 (0.29–0.38) & 0.18 (0.12–0.23) \\
                 & Kernel SHAP & 0.42 (0.38–0.47) & 0.28 (0.24–0.34) & $<0.01$ ($>-0.01$–$<0.01$) \\
                 & LIME & 0.42 (0.38–0.47) & 0.28 (0.25–0.33) & 0.01 ($<0.01$–0.01) \\
        \bottomrule
        \multicolumn{5}{l}{\footnotesize \textit{Note:} Values represent bootstrap-estimated Means (95\% BCa Confidence Interval).} \\
        \multicolumn{5}{l}{\footnotesize Values approaching zero are denoted using $<0.01$ and $>-0.01$ bounds.}
    \end{tabular}
    }

    \vspace{-20pt}
\end{table}

\subsection{Efficacy of Cross-Modal Attribution Projection} \label{subsec:attribution_source_impact}
IG emerged consistently as the optimal attribution method across all 3 evaluated attribution sources: derived from 12-lead ECG model, derived from CineECG model and 12-lead mapped to CineECG.
However, the post-processing requirements differed, with the mapped 12-lead projection achieving peak performance utilizing \emph{Scaled} normalization, while both direct CineECG and standard 12-lead attributions favored the \emph{Absolute} magnitude.

Table~\ref{tab:repr_1_global_comparison}  details the explanation alignment metrics comparing these native and mapped representations.
Projecting the 12-lead feature importance onto the spatial trajectory yielded a significantly higher mean Dice score of 0.56 compared to computing attributions directly on the CineECG model input, which scored 0.49, only marginally higher than the 12-lead baseline score of 0.47.
The most critical divergence appeared in the Spearman rank correlation analysis.
While the mapped projection and the raw baseline maintained positive rank correlations of 0.22 and 0.18 respectively, the direct CineECG attributions demonstrated a negligible rank alignment of 0.03.
This indicates that without the cross-modal mapping mechanism, the native 3D representation from the CineECG model fails to preserve the clinically relevant ranking of pathological features.

\begin{table}[ht!]

    \vspace{-10pt}
    \centering
    \caption{Impact of Attribution Source using the Integrated Gradients method.}
    \label{tab:repr_1_global_comparison}
    \resizebox{0.9\linewidth}{!}{
    \begin{tabular}{llccc}
        \toprule
        \textbf{Target Signal} & \textbf{Attribution} & \textbf{Dice Score} & \textbf{IoU Score} & \textbf{Spearman Cor.} \\
        \midrule
        CineECG Signal & CineECG Attr. & 0.49 (0.43–0.56) & 0.33 (0.29–0.41) & 0.03 (-0.07–0.12) \\
        CineECG Signal & 12-Lead (Mapped) & 0.56 (0.51–0.62) & 0.40 (0.35–0.47) & 0.22 (0.11–0.29) \\
        12-Lead Signal & 12-Lead Attr. & 0.47 (0.44–0.52) & 0.32 (0.29–0.37) & 0.18 (0.11–0.23) \\
        \bottomrule
        \multicolumn{5}{l}{\footnotesize \textit{Note:} Values represent bootstrap-estimated Means (95\% BCa Confidence Interval).}
    \end{tabular}
    }

    \vspace{-15pt}
\end{table}

\subsection{Sensitivity to Attribution Post-Processing} \label{subsec:post_processing_impact}
Finally, we evaluated the sensitivity to post-processing transformations by optimizing the experimental configurations exclusively across the available attribution methods.
Analysis of the winning configurations revealed that IG dominated as the optimal attribution technique across the majority of modality and post-processing combinations.
The sole exceptions were the \emph{Scaled} variant of direct CineECG attributions and the \emph{Positive} and \emph{Scaled} variants of the standard 12-lead signal, where Gradient SHAP achieved the highest Dice coefficient.

The aggregated results demonstrate a strong divergence in optimal post-processing strategies depending strictly on the source of the explanations.
Specifically, the mapped 12-lead configuration achieved its peak Dice coefficient utilizing \emph{Scaled} normalization, distinctively outperforming the alternative \emph{Absolute} and \emph{Positive} variants.
Conversely, the standard 12-lead attributions displayed a clear preference for the \emph{Absolute} magnitude, while direct CineECG attributions showed comparable Dice coefficient across multiple strategies.
Beyond Dice metric, a critical disparity emerged in the rank correlation metrics for the cross-modal mappings of 12-lead to CineECG with the \emph{Absolute} transformation causing the correlation to collapse below zero (-0.07).
Table~\ref{tab:detailed_3_prep_distribution} details these explanation alignment metrics for each evaluated post-processing variant.

\begin{table}[ht!]

    \vspace{-12pt}
    \centering
    \caption{Optimal Post-processing Strategy by Attributions Source.}
    \label{tab:detailed_3_prep_distribution}
    \resizebox{0.9\linewidth}{!}{
    \begin{tabular}{llccc}
        \toprule
        \textbf{Data Type} & \textbf{Prep Type} & \textbf{Dice Score} & \textbf{IoU Score} & \textbf{Spearman Cor.} \\
        \midrule
        CineECG  & Absolute & 0.49 (0.43–0.55) & 0.33 (0.29–0.41) & 0.03 (-0.08–0.12) \\
        Attributions                     & Positive & 0.47 (0.42–0.54) & 0.32 (0.27–0.39) & 0.01 (-0.03–0.07) \\
                             & Scaled   & 0.49 (0.43–0.56) & 0.34 (0.29–0.41) & 0.01 (-0.07–0.08) \\
        \midrule
        12-Lead  & Absolute & 0.47 (0.44–0.52) & 0.32 (0.29–0.37) & 0.18 (0.11–0.23) \\
        Attributions                     & Positive & 0.43 (0.40–0.48) & 0.28 (0.26–0.34) & 0.08 (0.06–0.10) \\
                             & Scaled   & 0.43 (0.40–0.48) & 0.28 (0.26–0.35) & 0.04 (0.03–0.06) \\
        \midrule
        12-Lead  & Absolute & 0.52 (0.48–0.58) & 0.36 (0.32–0.44) & -0.07 (-0.18–0.08) \\
        (Mapped)                 & Positive & 0.52 (0.47–0.59) & 0.37 (0.32–0.44) & 0.17 (0.08–0.25) \\
                         & Scaled   & 0.56 (0.51–0.62) & 0.40 (0.35–0.47) & 0.22 (0.11–0.29) \\
        \bottomrule
        \multicolumn{5}{l}{\footnotesize \textit{Note:} Values represent bootstrap-estimated Means (95\% BCa Confidence Interval).}
    \end{tabular}
    }
    \vspace{-15pt}
\end{table}

\subsection{Qualitative Evaluation of Explainability}
\begin{figure}[!bt]
    \vspace{-10pt}
  \centering
  \includegraphics[width=0.9\textwidth]{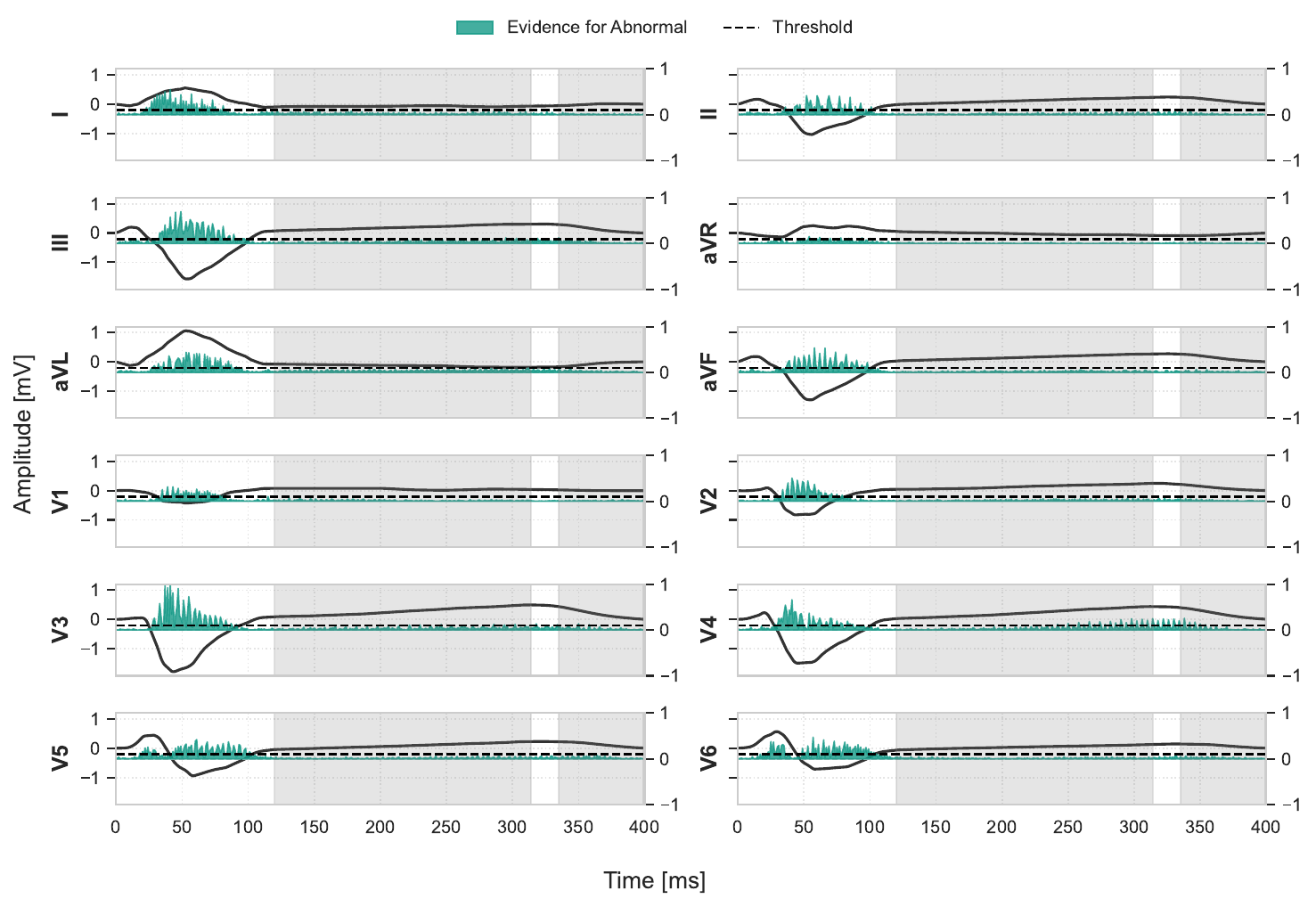}
  \caption{Expert annotation compared to model attributions (IG, absolute values) for case FID 18315. Model attributions marked with blue, IoU/Dice threshold marked with dotted line, expert annotations marked with white background.}
  \label{fig:expert_annotations}
\end{figure}

Case FID 18315 serves as a representative example of left anterior fascicular block (LAFB), where both the expert and the model demonstrate high diagnostic concordance.
The model predicted the abnormality with over 99\% certainty, aligning with the cardiologist's independent identification of clinical hallmarks, including left axis deviation and the characteristic qR morphology in lead aVL.
Despite spatial variations across individual leads, the attribution maps demonstrate strong alignment with expert temporal annotations, achieving a Dice coefficient of 0.71 and a Spearman correlation of 0.47.
Figure~\ref{fig:expert_annotations} illustrates that while the model correctly identifies the 30--100\,ms range as critical, the attribution magnitude is non-uniformly distributed, with peak intensities in leads III and V3 contrasting with negligible values in lead aVR.

\begin{figure}[bt!]
    \centering
    \begin{subfigure}[c]{0.45\textwidth}
        \centering
        \includegraphics[width=\textwidth]{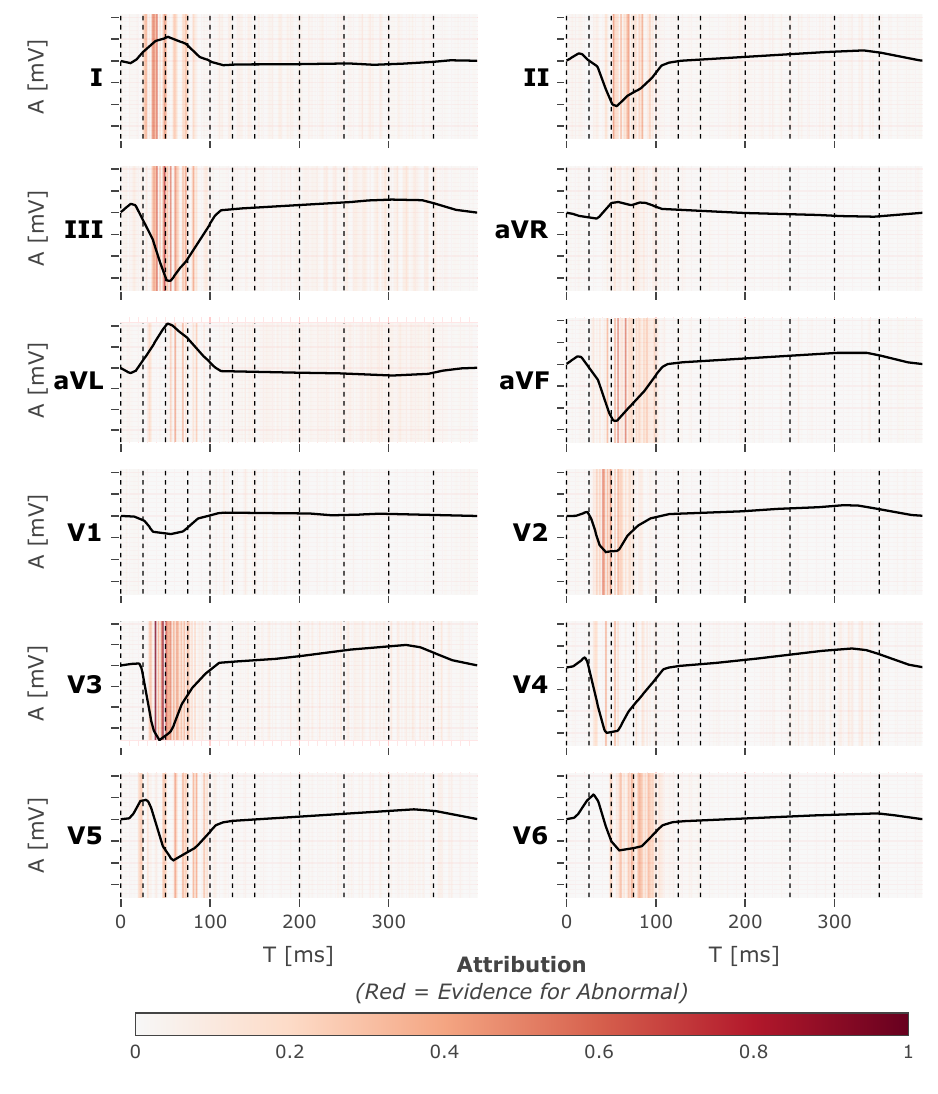}
        \caption{12-lead model attributions.}
        \label{fig:multimodal_explanation}
    \end{subfigure}\hspace{15pt}
    \begin{tabular}[c]{@{}c@{}}
        \begin{subfigure}[c]{0.3\textwidth}
            \centering
            \includegraphics[width=0.9\textwidth]{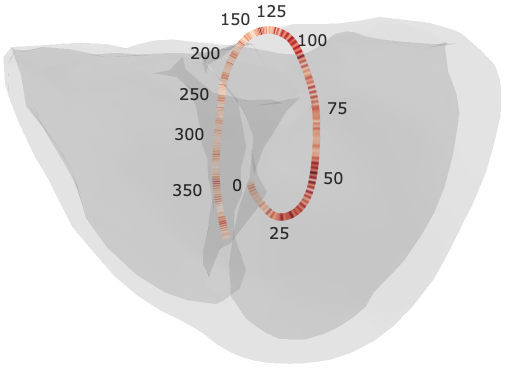}
            \caption{Direct CineECG Attributions}
            \label{fig:cine_direct}
        \end{subfigure}\\
        \noalign{\bigskip}
        \begin{subfigure}[c]{0.3\textwidth}
            \centering
            \includegraphics[width=0.9\textwidth]{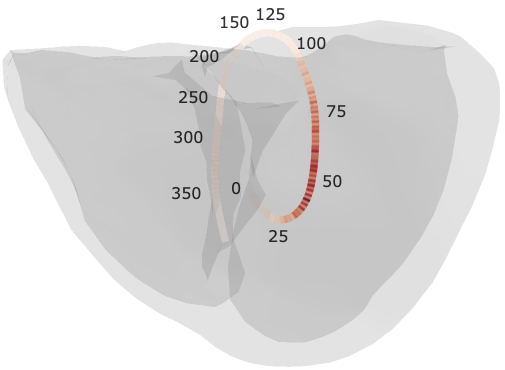}
            \caption{Mapped 12-Lead Attributions}
            \label{fig:cine_mapped}
        \end{subfigure}
    \end{tabular}
    \caption{Comparison of multimodal attributions for case FID 18315. 
    (a) 12-lead model attributions.
    (b) CineECG model attributions. 
    (c) 12-lead model attributions, projected onto the CineECG trajectory. 
    Attribution scale is the same for all figures. Numbers in (b) and (c) represent milliseconds.
    }
    \label{fig:cine_comparison}
    \vspace{-10pt}
\end{figure}

This highlights two systematic limitations of gradient-based feature attribution in the ECG domain.
First, importance values frequently collapse when the signal crosses the isoelectric line, a phenomenon visible in lead V5 and V6 where a distinct attribution dip occurs around 50\,ms as the signal approaches zero.
This suggests that vanishing gradients at zero-crossing points may impede the continuous estimation of feature importance.
Second, the model struggles to maintain high attribution magnitudes for temporally prolonged, low-amplitude features such as the T-wave.
As shown in the 325\,ms range, these clinically relevant segments often remain sub-threshold (with the minor exception of lead V4), indicating that the attribution mechanism is inherently biased toward high-frequency, high-amplitude morphological shifts rather than sustained potential changes.
Figure~\ref{fig:multimodal_explanation} further visualizes these patterns across the multimodal representations, providing a basis for comparing native and projected explanations.

\subsubsection{CineECG mapping}
We visualize the classifier attributions mapped to the CineECG domain as 3D renders, overlaying a semi-transparent ventricular model to provide anatomical context.
Figure~\ref{fig:cine_comparison} contrasts the attributions derived directly from the CineECG model against those mapped from the 12-lead baseline for the LAFB case (FID 18315).
The direct CineECG explanation (Fig.~\ref{fig:cine_direct}) displays an intense focus solely on the signal onset (0–25 ms), ignoring the subsequent morphology of the cardiac cycle.
In contrast, the mapped attribution (Fig.~\ref{fig:cine_mapped}) correctly ignores the non-pathological onset, concentrating the importance along the trajectory segment corresponding to ventricular depolarization (approx. 50–100 ms).
This alignment confirms that the proposed mapping successfully transfers the clinically relevant localization of the fascicular block from the 12-lead domain into the 3D space.

\section{Discussion} \label{sec:discussion}
Expert evaluation confirms that projecting 12-lead ECG attributions into 3D space highlights the clinical utility of CineECG.
Relying exclusively on the spatial domain for training yields lower accuracy and incoherent explanations, evidenced by near-zero Spearman correlations.
This necessitates using the standard 12-lead ECG as a robust feature extractor.
Mapping attributions onto the 3D trajectory achieved a 0.56 Dice coefficient, surpassing the 0.47 for 12-lead baseline. 
This demonstrates that the true value of the 3D representation lies in its capacity to contextualize temporal features rather than replacing them.

The success of this cross-modal mapping depends heavily on the underlying interpretability mechanism.
IG combined with \emph{Scaled} post-processing provided the most effective explanations by preserving the continuous spatial topology required for anatomical projection.
Conversely, perturbation-based approaches like LIME and Kernel SHAP proved inadequate for this domain.
These methods failed to maintain meaningful rank correlations on the high-dimensional signal, rendering them unsuitable for precise spatial localization.

Spatial alignment quality remains intrinsically linked to the clinically salient characteristics of the recorded cardiac activity.
Alignment scores were consistently higher for pathological cases, where focal anomalies like ST-segment deviations facilitate precise localization by both models and human experts.
Confirming a normal rhythm, conversely, requires a wider inspection of the entire cardiac cycle.
This global assessment inherently limits potential spatial overlap with the highly localized attribution maps generated by the network.

A critical finding emerged regarding the behavior of mapped attributions during classification failures.
The spatial projection achieved its best alignment with human experts precisely when the final model prediction diverged from the established diagnosis.
This paradox indicates that the attribution mechanism successfully localizes correct pathological morphology despite an incorrect label from the classification head.
Such behavior establishes the cross-modal mapping as a valuable diagnostic debugging tool, empowering clinicians to verify network focus when the automated system fails.

Aggregating feature attributions onto the 3D representation reduces the cognitive load of interpreting twelve separate temporal plots and allows clinicians to validate decisions against known anatomical activation patterns rather than abstract morphological notches.
The qualitative analysis of the LAFB case explicitly demonstrates this advantage.
Explanations derived directly from the native CineECG model erroneously fixated on the non-pathological signal onset.
Conversely, the cross-modal mapping successfully transferred the clinically relevant focus to the ventricular depolarization segment.

Furthermore, the spatial integration of CineECG acts as a natural filter for the inherent limitations of gradient-based methods.
Temporal attributions frequently collapse at the isoelectric line or fade across prolonged, low-amplitude features like the T-wave.
Mapping these unstable gradients onto a continuous 3D trajectory masks these inconsistencies, presenting the clinician with a cohesive pathological region.
Projecting attributions from the robust 12-lead architecture onto the anatomical model circumvents the traditional trade-off between predictive performance and interpretability.
This cross-modal mechanism allows the spatial representation to inherit the diagnostic rigor of the standard model while providing intuitive clinical contextualization.

Despite the demonstrated clinical utility of the proposed framework, certain limitations must be acknowledged.
The necessity for full differentiability to compute stable gradient-based attributions constrained our architecture to a standardized 1D-ResNet with architectural solutions allowing for better model attributions.
Consequently, we opted against employing highly complex state-of-the-art ensembles, although the selected model maintained robust diagnostic accuracy.
Furthermore, while the evaluation on 20 expert-annotated cases establishes a strong proof-of-concept for cross-modal mapping, future validation across larger clinical cohorts is warranted.

\section{Conclusions} \label{sec:conclusions}
We demonstrated that the explainability of deep learning models can be significantly enhanced by mapping 12-lead feature attributions onto the CineECG anatomical trajectory.
This cross-modal mapping yields superior alignment with expert ground truth compared to both raw temporal attributions and models trained directly on spatial data.
Furthermore, through averaging the 12-lead attribution, the mapping mechanism functions as a spatial regularizer that filters attribution noise while preserving the salient localization of pathology.
Mapping temporal feature attributions to anatomical structures enables intuitive clinical verification, reconciling diagnostic accuracy with explanatory clarity.

\begin{credits}
\subsubsection{\ackname}This paper is part of a project that has received funding from the European Union's Horizon Europe Research and Innovation Programme, under Grant Agreement number 101120406. 
The paper reflects only the authors' view, and the EC is not responsible for any use that may be made of the information it contains. 

The contribution of Karol Dobiczek and Maciej Mozolewski for the research for this publication has been additionally supported by a grant from the Priority Research Area (DigiWorld) under the Strategic Programme Excellence Initiative at Jagiellonian University.

The authors thank the Faculty of Physics, Astronomy and Applied Computer Science at Jagiellonian University for computational resources.

During the preparation of this manuscript, the authors used Google Gemini Pro as an assistive tool for: 
(i) language editing and minor \LaTeX\ formatting, 
(ii) proofreading, 
(iii) literature search support limited to shortlisting (all cited sources were independently verified from the originals), and 
(iv) coding assistance for computational tasks. 
All of these outputs were reviewed, tested, and validated by the authors. 
The tool did not generate any datasets, experimental results, figures, or conclusions. 
No confidential or personal data were provided to the system. 
The authors retain full responsibility for the content and integrity of this article. 
\end{credits}
\bibliography{references}{}
\bibliographystyle{plain}

\end{document}